\documentclass[
superscriptaddress,
amsmath,amssymb,
aps,
pra,
longbibliography
]{revtex4-2}

\usepackage{graphicx}
\usepackage{dcolumn}
\usepackage{bm}

\begin{document}


\title{Critical Power for Temporal-Pulse Collapse in Third-Harmonic Generation}

\author{Wataru Komatsubara}
\affiliation{%
 Department of Physics, The University of Tokyo, 7-3-1 Hongo, Bunkyoku, Tokyo, Japan
}%
\author{Kuniaki Konishi}%
\affiliation{%
 Institute for Photon Science and Technology, The University of Tokyo, 7-3-1 Hongo, Bunkyoku, Tokyo, Japan
}%
\author{Junji Yumoto}
\author{Makoto Kuwata-Gonokami}
 \email{gonokami@phys.s.u-tokyo.ac.jp}
\affiliation{%
 Department of Physics, The University of Tokyo, 7-3-1 Hongo, Bunkyoku, Tokyo, Japan
}%
\affiliation{%
 Institute for Photon Science and Technology, The University of Tokyo, 7-3-1 Hongo, Bunkyoku, Tokyo, Japan
}%

\date{\today}

\begin{abstract}
The self-trapping critical power of light propagation is one of the key physical quantities characterizing nonlinear-beam propagation. Above the critical power, the spatial and temporal profiles of the beam deviate from its original shapes. Therefore, the critical power is considered an important indicator in nonlinear optical phenomena, such as filamentation and laser processing. However, although the concept of the critical power has been established for fundamental waves, it remains unclear if the power-dependent phenomena can also be observed in harmonic generation because of the complex interplay of nonlinear-propagation effects and ionization-plasma effects. In this study, we find the critical power for the third-harmonic generation; the criterion for whether or not temporal-pulse collapse occurs in the third-harmonic generation is determined only by the incident power. Experiments show that a certain incident power exists, at which the third-harmonic power exerts a specific dependence, independent of the focusing conditions. This incident power is approximately six times lower than the self-trapping critical power of the fundamental pulse, which indicates that it is unique to the third-harmonic generation. Numerical calculations reveal that at this incident power, the third-harmonic pulse begins to collapse temporally. Furthermore, the numerical calculations reproduce the experimental results without the nonlinear effects of the fundamental pulse, dispersion effect, and ionization-plasma effects. This shows that the pulse collapse is due to the interference effects from the third-order nonlinear term, which disappears after long-distance propagation due to phase mismatch, and higher-order nonlinear terms. This study demonstrates the existence of the critical power for harmonic pulses and show that higher-order nonlinear effects on the harmonics can yield a universal phenomenon. By extracting and organizing the underlying physics from the complex interplay of the various nonlinear processes, these results provide important insights for the exploration and control of higher-order nonlinear effects on harmonics.
\end{abstract}

\maketitle

\section{INTRODUCTION}
Recent advances in intense, ultrashort-pulse technology have significantly improved the controllability of nonlinear phenomena \cite{strickland1985compression}. Consequently, it is now possible to finely elucidate the mechanisms of complex nonlinear phenomena, where various processes are intricately intertwined. In nonlinear phenomena occurring in air, self-focusing due to the third-order nonlinear effects  \cite{marburger1975self}, diffraction due to laser propagation, and plasma generation are complexly interconnected  \cite{COUAIRON200747, doi:10.1139/p05-048}. These nonlinear phenomena in air are exploited in filamentation  \cite{theberge2006tunable, chin2007filamentation, theberge2007self} and THz generation  \cite{roskos2007broadband}. The nonlinear effects of air also modulate the spatial and temporal profiles of lasers, evidencing their essential role in laser processing \cite{pasquier2016predictable}.

One of the important physical parameters for characterizing nonlinear effects is the self-trapping critical power, at which the effects of self-focusing and beam diffraction are balanced \cite{boyd2020nonlinear, couairon2003dynamics}. The critical power is independent of the focusing conditions and dependent only on the incident power \cite{boyd2020nonlinear}. If the incident power is above the critical power, the beam profile of the fundamental wave will not maintain its original shape. This concept of the critical power has been extended to the time domain with the advent of intense ultrashort-pulse lasers \cite{rothenberg1992pulse, ranka1996observation, chernev1992self, rothenberg1992space, fibich1997self, chernev1992self}. When the incident power exceeds the critical power, the time profile of the fundamental pulse also does not maintain its original shape.

The critical power of the fundamental wave has also been employed as an indicator to characterize the complex nonlinear dynamics in harmonic generation. For example, the spatial profile of the Third-Harmonic (TH) wave adopts a ring structure when the fundamental wave is irradiated at power  levels above the critical power \cite{theberge2005third, theberge2007conical, steponkevivcius2014spectral}. When the fundamental wave is irradiated at power levels higher than the critical power, the spatial profile of the fundamental wave adopts a different shape from its original shape, which induces the modulation of the spatial profile of the TH wave.

However, it is not obvious that the concept of the critical power is unique to the fundamental wave. This is because the critical power of the fundamental pulse is determined based on the third-order nonlinear effects, although the lowest-order nonlinear terms of the TH disappear because of phase mismatch, and the higher-order nonlinear terms are observed \cite{boyd2020nonlinear, liu2011efficient}. Therefore, considering that the nonlinear effects in the TH propagation are different from those of the fundamental pulse, a ``critical power” unique to the TH wave may exist, i.e., the phenomenon unique to the TH generation, dependent only on the incident power of the fundamental wave and  independent of the focusing condition.

In addition to the propagation effects described above, the TH generation is also influenced by ionization and plasma generation effects, in combination with the nonlinear effects of the fundamental wave \cite{akozbek2002third, akozbek2003continuum}. Due to the complex interplay of these various effects \cite{moll2002conical, vaivcaitis2009conical, bejot2010higher, kolesik2010higher, kolesik2010femtosecond, bejot2011transition, kosareva2011arrest, bejot2011higher, nath2013seventh, weerawarne2015higher}, no discussion has been tabled on the existence of the critical power specific to the TH pulses. However, considering that ionization and plasma generation are intensity-dependent \cite{boyd2020nonlinear}, whereas the critical power is power-dependent, the different focusing conditions should be employed to separate the complex nonlinear effects.

In this study, we demonstrate that the threshold for whether or not temporal-pulse collapse in the third-harmonic generation occurs is determined only by the incident power, as the critical power of the fundamental wave. First, we show that the power dependence of the TH generation in air varies depending on the focusing conditions, and its experimental results are reproduced by numerical simulations. Detailed numerical simulations reveal the existence of a certain incident power that yields a specific-power dependence of the TH pulse. This power is independent of the focusing conditions and is as small as 1/6 of the critical power for the fundamental wave. Numerical calculations of the TH-pulse profiles reveal that above this incident power, the time profile of the TH pulse does not maintain its original shape, which demonstrates this specific power-dependent behavior. Furthermore, we verify that this temporal collapse of the TH pulse still occurs without the nonlinear effects of the fundamental pulse and dispersion, as well as the effects of air ionization and plasma. We show that the temporal collapse is attributed to only the nonlinear effect of the TH pulse and its propagation. In addition, we confirm that the change in the spatial profile of the TH pulse is irrelevant to this incident power. Therefore, this incident power, which is lower than the critical power of the fundamental wave, can be denoted as the``time-collapse critical power,” specific to the TH pulse.

\section{EXPERIMENT}

\begin{figure}
\centering
\includegraphics[width=86mm]{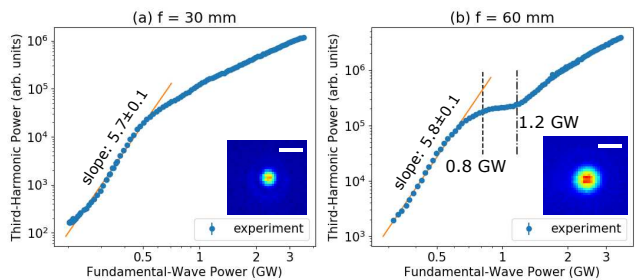}
\caption{\label{fig1} Experimental TH-power dependences using (a) f = 30 mm and (b) f = 60 mm focusing lenses and beam profiles (insets) measured at the focal point. The white bars in the insets are 10 $\mu$m.}
\end{figure}

The TH is generated in air under different focusing conditions, and the TH-power dependence is measured. Even when the incident power is the same, the intensity of the fundamental wave differs depending on the focusing conditions, which can lead to different TH-power dependences. A Ti:sapphire laser source (center wavelength: 800 nm, pulse width: 45 fs, repetition rate: 1 kHz) was used as the light source. For details on the experiment, see Appendix \ref{AppA}. Figures \ref{fig1}(a) and \ref{fig1}(b) show the TH-power dependences with lenses having focusing distances f of 30 and 60 mm, and the measured focusing diameters are 4.4 and 7.2 $\mu$m, respectively. The results show that the TH-power dependences behave differently depending on the focusing conditions. For the case in Figure \ref{fig1}(a), the TH power monotonically increases at a power of 5.7 initially and saturates around 0.5 GW. For the case in Figure \ref{fig1}(b), the TH power monotonically increases at the power of 5.8 in the low-fundamental-power region, after which it saturates at 0.8 GW and subsequently increases again at 1.2 GW. The experimental result in Figure \ref{fig1}(a) is consistent with the results of previous studies \cite{ganeev1986optical, l1988third, fedotov1997saturation, marcus1999third, ganeev2000harmonic, chang2001third, ganeev2006third, malcuit1990anomalies}, in which the TH-power dependence increases to the 5th power instead of the 3rd power. This occurs because the lowest-order nonlinear term, which follows the third power of the fundamental pulse, disappears because of phase mismatch. The experimental result shown in Figure \ref{fig1}(b) is, however, different from that in Figure \ref{fig1}(a), and we observe that saturation occurs at 0.8 GW. As mentioned before, considering that ionization and plasma generation are intensity-dependent, whereas the critical power is power-dependent, the different focusing conditions could lead to the different power dependences. These experimental results strengthen this consideration, and suggest that the mechanism of the TH generation varies with the focusing condition.

\section{SIMULATIONS AND DISCUSSION}
To clarify the differences in the TH-power dependences, numerical calculations were performed that incorporated the effects of absorption and scattering by plasma and multiphoton ionization, in addition to third-order Kerr nonlinear effects. The equations are as follows \cite{liu2011efficient, ganeev2006third, akozbek2002third, akozbek2003continuum, theberge2005third, theberge2007conical}:

\begin{align}
    & \left\{ \nabla_{\perp}^2 + 2ik_{\omega} \frac{\partial}{\partial z} + 2k_{\omega} \left( ik'_{\omega} \frac{\partial}{\partial t} - \frac{k''_{\omega}}{2}\frac{\partial^2}{\partial t^2} \right) \right\} A_1 (\bm{r}, t) \nonumber \\
    & = \sum_{p=O_2, N_2} (N_p - N_{ep}) / N_{mol}\times \left[ - \frac{3\omega_0^2 \chi^{(3)}}{c^2} \times \right. \nonumber \\
    &\left. \left\{A_1^2 A_1^{\ast} + 2 A_1 A_3 A_3^{\ast} + (A_1^{\ast})^2 A_3 \exp \left(i ( k_{3\omega} - 3k_{\omega}) z \right)  \right\} \right. \nonumber \\
    & \left. - i k_{\omega} \beta_{p, \omega} |A_1|^{2 K_{p, \omega} - 2} A_1 \right]- i k_{\omega} \sigma_{\omega} (1 + i \omega \tau_c) \rho A_1, \label{eq1}
\end{align}
\begin{align}
    & \left\{ \nabla_{\perp}^2 + 2ik_{3\omega} \frac{\partial}{\partial z} + 2k_{3\omega} \left( ik'_{3\omega} \frac{\partial}{\partial t} - \frac{k''_{3\omega}}{2}\frac{\partial^2}{\partial t^2} \right) \right\} A_3 (\bm{r}, t) \nonumber \\
    &= \sum_{p=O_2, N_2} (N_p - N_{ep}) / N_{mol}\times \left[ - \frac{(3\omega_0)^2 \chi^{(3)}}{c^2} \times \right. \nonumber \\
    & \left. \left\{ 6 A_3 A_1 A_1^{\ast} + 3 A_3^2 A_3^{\ast} + A_1^3 \exp \left(i (3k_{\omega} - k_{3\omega}) z \right)  \right\} \right. \nonumber \\
    &\left. -i k_{3\omega} \beta_{p, 3\omega} |A_3|^{2K_{p, 3\omega} - 2} A_3 \right] -i k_{3\omega} \sigma_{3\omega} (1 + i 3\omega \tau_c) \rho A_3, \label{eq2}
\end{align}
\begin{align}
    \rho = \sum_{p=O_2, N_2} N_{ep}, \label{eq3}
\end{align}
\begin{align}
    & \frac{\partial N_{ep}}{\partial t} = (N_p - N_{ep})(\sigma_{K_{p, \omega}} |A_1|^{2 K_{p, \omega}} + \sigma_{K_{p, 3\omega}} |A_3|^{2 K_{p, 3\omega}}) \nonumber \\
    & + (\sigma_{\omega} \frac{|A_1|^2}{U_p} + \sigma_{3\omega} \frac{|A_3|^2}{U_p} )N_{ep} (p=O_2, N_2), \label{eq4}
\end{align}
where $A_1,A_3$ are the envelope functions; $k_{\omega},k_{3\omega}$ are the wave-number vectors; $k'_{\omega},k'_{3\omega}$ are the inverses of the group velocity; $k''_{\omega},k''_{3\omega}$ are the group-velocity dispersions of the fundamental and TH pulses, respectively; $\omega$ is the angular frequency of the fundamental pulse; $c$ is the speed of light; $\chi^{(3)}$ is the third-order nonlinear susceptibility; $\Delta k=3k_{\omega}-k_{3\omega}$; $\tau_c$ is the collision time of the electron; $N_{mol}$ is the density of neutral molecules; $N_p,N_{ep}$ are the densities of molecules and electrons, respectively; $\sigma_{K_{p,\omega}},\sigma_{K_{p,3\omega}}(p=O_2,N_2)$ are the ionization cross-sections of the fundamental and TH pulses, respectively; $\beta_{p,\omega}=K_{p,\omega} \hbar \omega \sigma_{K_{p,\omega}} \rho_{nt}$; $\beta_{p,3\omega}=K_{p,3\omega} \hbar 3\omega \sigma_{K_{p,3\omega}} \rho_{nt}$; $\rho_{nt}$ is the density of neutral molecules; $\sigma_{\omega}=e^2/(\varepsilon_0 c^2 m_e k_{\omega} \omega \tau_c ),$ and $\sigma_{3\omega}=e^2/(\varepsilon_0 c^2 m_e k_{3\omega} 3\omega \tau_c )$. The initial spatial and temporal profiles of the fundamental pulse is assumed to be Gaussian shapes. (See Appendix \ref{AppB} for the details).

\begin{figure*}
\includegraphics[width = 170mm]{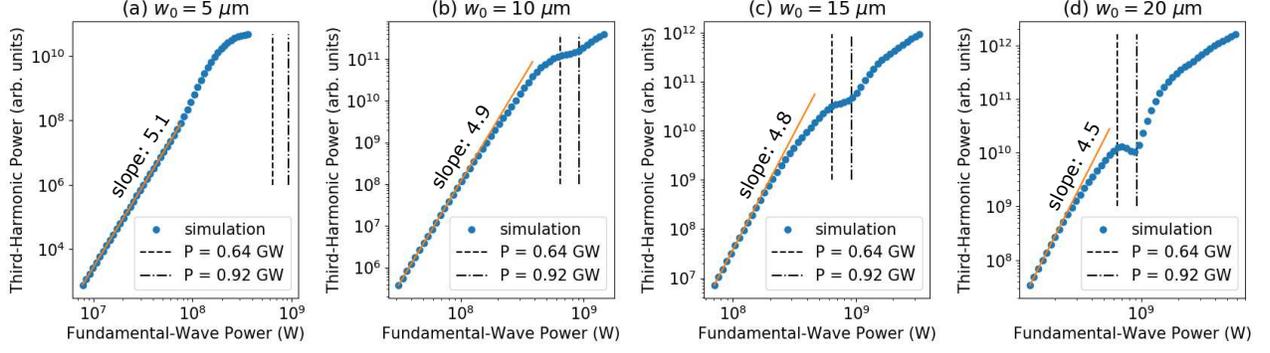}
\caption{\label{fig2} Calculated TH-power dependences with the beam waists: (a) $w_0$ = 5 $\mu$m, (b) $w_0$ = 10 $\mu$m, (c) $w_0$ = 15 $\mu$m, and (d) $w_0$ = 20 $\mu$m.}
\end{figure*}

The numerical results are shown in Figure \ref{fig2}. As shown in Figure \ref{fig2}(a), when the focusing diameter is $w_0$ = 5 $\mu$m, the calculated TH-power dependence reproduces an initial monotonic increase, followed by saturation, as shown in Figure \ref{fig1}(a). Preferably, for the case in Figure \ref{fig2}(b) ($w_0$ = 10 $\mu$m), a monotonic increasing trend is observed at the low-fundamental-power region; however, the TH power saturates at an incident power of 0.64 GW. After that, the TH power increases to around 0.92 GW and finally achieves saturation. This result corresponds to the experimental result in Figure \ref{fig1}(b). In Figure \ref{fig2}(b), the incident power of 0.64 GW, at which the specific structure appears, is in quantitative agreement with the incident power of 0.8 GW [Fig. \ref{fig1}(b)]. The experimental results [Figs. \ref{fig1}(a,b)] are, therefore, reproduced by numerical calculation [Figs. \ref{fig2}(a,b)].

Figures \ref{fig2}(c,d) show the results of the calculations with relatively large focusing diameters. In Figures \ref{fig2}(c) ($w_0$ = 15 $\mu$m) and (d) ($w_0$ = 20 $\mu$m), the specific-TH-power dependences also appear at the incident power of 0.64 GW, after which they begin to increase again from 0.92 GW, similar to the result in Figure \ref{fig2}(b). Furthermore, the power at which this specific-TH-power dependence begins to appear, 0.64 GW, is independent of the focusing diameter. The numerical calculations in Figure \ref{fig2} show that the power dependence profile of the TH pulse adopts different shapes depending on the focusing conditions. This also suggests that there is a mechanism that relies on the incident power with large focusing diameters.

Notably, the incident power at which the TH-power dependence begins to display a specific structure in Figures \ref{fig2}(b, c, d) is 0.64 GW, which is independent of the focusing diameter. This independence from the focusing diameter is similar to the case with the self-trapping critical power of the fundamental pulse. Furthermore, the intensity independence suggests that this incident power is not relevant to ionization. Figure \ref{fig2}(a) shows that the TH-power dependence already reaches saturation at power levels less than 0.64 GW. This is presumably due to the ionization effect and because even at the same incident power, the intensity increases as the focusing diameter decreases. Here the critical power of the fundamental pulse is estimated as follows \cite{boyd2020nonlinear}:

\begin{align}
    P_{cr} = \frac{\pi (0.61)^2 \lambda^2}{8n_0 n_2} = 3.7 \; \mathrm{GW}. \label{eq5}
\end{align}

The above power of 0.64 GW is approximately 1/6 of the fundamental critical power, suggesting that this power is unique to the TH generation. In fact, at 0.64 GW, which is lower than the critical power of the fundamental pulse, the temporal and spatial profiles of the fundamental pulse maintain their original shapes (see Appendix \ref{AppC} for details).

\begin{figure}
\centering
\includegraphics[width=86mm]{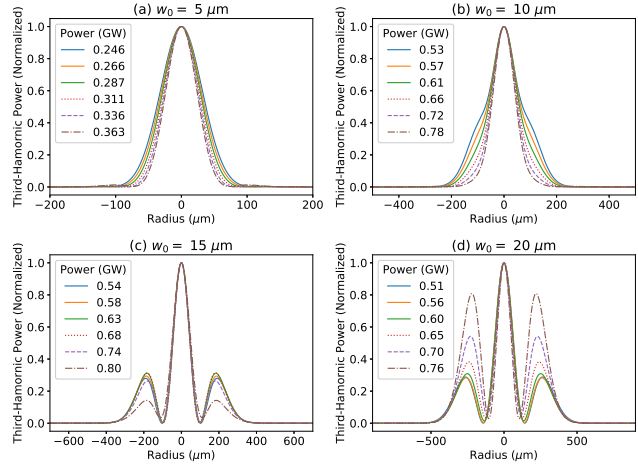}
\caption{\label{fig3} Calculated TH-spatial profiles at $t=0$, with the beam waists: (a) $w_0$ = 5 $\mu$m, (b) $w_0$ = 10 $\mu$m, (c) $w_0$ = 15 $\mu$m, and (d) $w_0$ = 20 $\mu$m.}
\end{figure}

The power dependence of the TH pulse results in the loss of information regarding the pulse profiles because the results are obtained by spatial and temporal integration. To clarify the spatial- and temporal-profile characteristics of the TH pulse, additional numerical calculations are conducted. The spatial profile of the TH pulse around 0.64 GW is illustrated in Figure 3. For $w_0$ = 5, 10 $\mu$m in Figures 3(a, b), the spatial profile maintains its original shapes, although it exhibits a slight distortion as the incident power increases. Contrarily, as shown in Figures 3(c, d) for $w_0$ = 10, 15 $\mu$m, the spatial profile is already split, deviating significantly from its original shape. This is observed even at incident powers below 0.64 GW and does not reflect the TH-power-dependent specific behavior, i.e., the splitting of the spatial profile for the TH pulse is irrelevant to the incident power.

\begin{figure}
\centering
\includegraphics[width=86mm]{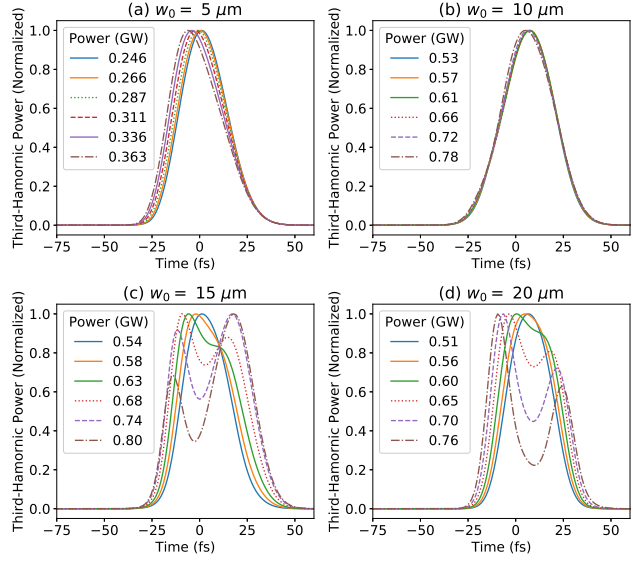}
\caption{\label{fig4} Calculated TH-temporal profiles at $r=0$, with the beam waists: (a) $w_0$ = 5 $\mu$m, (b) $w_0$ = 10 $\mu$m, (c) $w_0$ = 15 $\mu$m, and (d) $w_0$ = 20 $\mu$m.}
\end{figure}

The time profiles of the TH pulse around 0.64 GW are shown in Figure 4. As shown in Figures 4(a, b) for $w_0$ = 5, 10 $\mu$m, the time profiles do not change with the incident power and maintain their original shapes. Conversely, for $w_0$ = 15, 20 $\mu$m, as shown in Figures 4(c, d), the pulse-profile shape changes from the original shape when the incident power exceeds 0.6 GW. The incident power of 0.6 GW is almost equal to the incident power of 0.64 GW suggested in Figure 2. Therefore, the physical origin of the specific-TH-power-dependent structure may be the splitting of the TH pulse in the time domain. Further, this incident power of 0.64 GW can be regarded as the ``critical power” for the TH pulse, at which the time profile begins to deviate from its initial shape.

To investigate the physical origin of the temporal splitting for the TH pulse at 0.64 GW, we perform numerical calculations of the TH generation, neglecting the dispersion, ionization-plasma effects, and third-order nonlinear effects of the fundamental pulse. When the third-order nonlinear effects of the fundamental wave are neglected, the solution for the fundamental pulse is the same as that for propagating in a vacuum and written as

\begin{align}
    A_1 (r,z) = \frac{E_1}{1 + i \zeta} \exp \left\{ - \frac{r^2}{w_0^2 (1 + i \zeta) }\right\} \exp \left\{ - \frac{t^2}{\tau^2}\right\}, \label{eq6}
\end{align}
where $E_1$ is the electric field, and $w_0$ is the beam waist of the fundamental pulse; $\zeta =z/z_R$, $z_R$ is the rayleigh length, ant $\tau$ is the pulse width. The nonlinear-propagation equation for the TH pulse is given by

\begin{align}
    \frac{\partial A_3}{\partial z} & = \frac{i}{2k_3} \nabla_{\perp}^2 A_3 + \nonumber \\
    & i \frac{(3\omega)^2}{2k_3 c^2} \chi^{(3)} \left\{ A_1^3 \exp(i \Delta k z) + 6 |A_1|^2 A_3  \right\} . \label{eq7}
\end{align}

The solution for Eq. (\ref{eq7}) without the second nonlinear term, $6|A_1|^2 A_3$, is obtained by substituting Eq. (\ref{eq6}) into Eq. (\ref{eq7}), as follows \cite{boyd2020nonlinear}:

\begin{align}
    & A_3 (r,z) = \nonumber \\
    & \frac{- \frac{3\omega^2}{2k_1^2c^2}\chi^{(3)}E_1^3 (k_1w_0)^2}{2(1+i\zeta)} \exp\left\{ -\frac{3r^2}{w_0^2 (1+i\zeta)}\right\} \exp \left\{ -\frac{3t^2}{\tau^2} \right\}, \label{eq8}
\end{align}
where $\Delta k$ is approximated to 0. $A_3$ vanishes because of phase mismatch, and the Gaussian shape is maintained in both space and time domains. If we include the second term, $6|A_1|^2 A_3$, Eq. (\ref{eq7}) can be converted into

\begin{align}
    & \frac{\partial A_3}{\partial z} = \frac{i}{2k_3} \nabla_{\perp}^2 A_3 + i \frac{(3\omega)^2}{2k_3 c^2} \chi^{(3)} \times \left\{ \frac{E_1^3}{(1+i\zeta)^3} \right. \nonumber \\
    & \left. \times \exp \left\{ -\frac{3r^2}{w_0^2 (1+i\zeta)} \right\} \exp \left\{ -\frac{3t^2}{\tau^2}\right\} \exp(i\Delta k z)\right. \nonumber \\
    & \left. + \frac{6E_1^2}{1+\zeta^2} \exp\left\{ -\frac{2r^2}{w_0^2 (1+\zeta^2)} \right\} \exp\left\{ -\frac{2t^2}{\tau^2} \right\}A_3 \right\} . \label{eq9}
\end{align}
The second and third terms on the right-hand side of Eq. (\ref{eq9}) have different exponential functions with respect to space and time; therefore, the solution is not Gaussian and cannot be solved analytically.

\begin{figure}
\centering
\includegraphics[width=86mm]{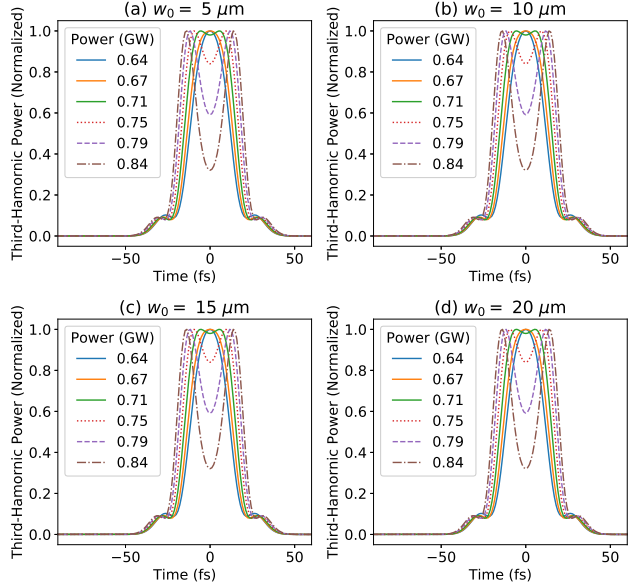}
\caption{\label{fig5} TH temporal profiles at $r=0$, calculated according to Eq. (\ref{eq5}), with the beam waists: (a) $w_0$ = 5 $\mu$m, (b) $w_0$ = 10 $\mu$m, (c) $w_0$ = 15 $\mu$m, and (d) $w_0$ = 20 $\mu$m.}
\end{figure}

The numerical calculation results for Eq. (\ref{eq9}) are shown in Figure \ref{fig5}. ($\Delta k$ is approximated to 0.) Figure \ref{fig5} shows that the pulse shape in the time domain is the same regardless of $w_0$ and that the pulse begins to split when the incident power exceeds 0.7 GW. The temporal splitting of the TH pulse is reproduced by solving only Eq. (\ref{eq9}), which ignores the third-order nonlinear effects of the fundamental pulse in Eq. (\ref{eq6}). This strongly supports that the critical power found in this study is independent of the fundamental-wave critical power and unique to the TH pulse.

Different from the case of $w_0$ = 5, 10 $\mu$m in Figures \ref{fig4}(a, b), where the pulses do not collapse, they do in the case in Figures \ref{fig5}(a, b). This difference may be due to the neglect of the ionization-plasma effects. In fact, at relatively small $w_0$ values, the ionization-plasma effects cannot be negligible since the intensity increases at the same incident power. Furthermore, the results in Figure \ref{fig5} are the same independent of $w_0$, showing consistency with the case where the ionization-plasma effects are ignored.

\begin{figure}
\centering
\includegraphics[width=86mm]{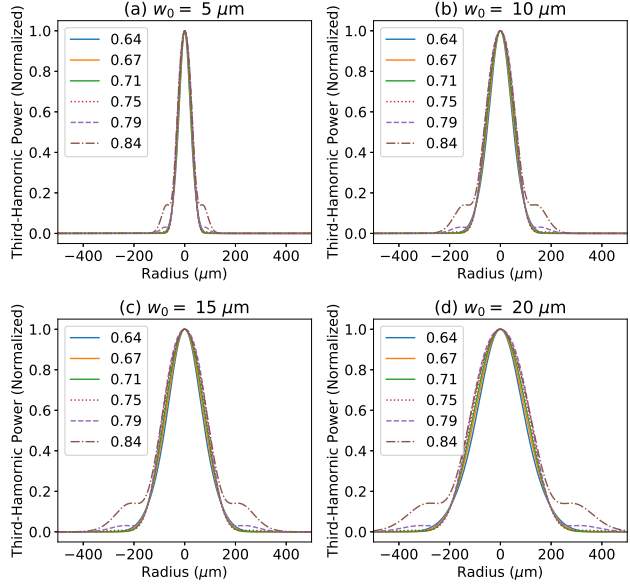}
\caption{\label{fig6} TH spatial profiles at $t=0$, calculated according to Eq. (\ref{eq5}), with the beam waists: (a) $w_0$ = 5 $\mu$m, (b) $w_0$ = 10 $\mu$m, (c) $w_0$ = 15 $\mu$m, and (d) $w_0$ = 20 $\mu$m.}
\end{figure}

The calculated results for the spatial profile are shown in Figure \ref{fig6}. Figure \ref{fig6} shows that the pulses do not collapse and that they have the same shape, regardless of the focusing diameter. These numerical results reinforce the discussion of Figure \ref{fig3}(c, d), that the behavior of the spatially collapsing pulse is not dependent on the power. Therefore, the reason why the pulses in Figures \ref{fig3}(c, d) are spatially broken is the nonlinear effects of the fundamental wave and the ionization-plasma effects, not the critical power for the TH pulse. The numerical results in Figure \ref{fig6} also show that the power that exhibits a specific-power dependence for the TH pulse collapses the temporal profile and not the spatial profile.

The critical power for the TH pulse discovered in this study is similar to that for the fundamental pulse, a phenomenon independent of the focusing diameter and dependent on the incident power. As another property, the critical power of the fundamental pulse is proportional to the inverse of the nonlinear susceptibility, as expressed in Eq. (\ref{eq5}). The similar property is confirmed for the critical power of the TH pulse. When numerical calculations are performed with the nonlinear susceptibility doubled, the critical power at which the specific-TH-power-dependence structure begins to appear is halved. This result is similar to that for the critical power of the fundamental pulse (see Appendix \ref{AppD} for details).

The temporal collapse for the TH pulse is reproduced even if the dispersion effects are ignored ($k'_{\omega},k'_{3\omega},k''_{\omega},k''_{3\omega}$ in Eq. (\ref{eq1}, \ref{eq2}) are all set to 0), which shows that the temporal collapse is unrelated to the effect of dispersion and considered to be different from the phenomena caused by dispersion-dependent nonlinear effects, such as self-steepening \cite{boyd2020nonlinear, ranka1998breakdown}. Therefore, the pulse collapse in the time domain can be attributed to the interference effect of the nonlinear terms of the TH: the second term on the right-hand side of Eq. (\ref{eq9}) for the third-order nonlinear polarization $\exp \left\{ -\frac{3t^2}{\tau^2} \right\}$, which is canceled by the phase mismatch, and the third term of right-hand side of Eq. (\ref{eq3}) for the higher-order term $\exp\left\{ -\frac{2t^2}{\tau^2}\right\} A_3$.

\section{CONCLUSION AND OUTLOOK}
We demonstrate that the criterion for whether temporal-pulse collapse in the TH generation occurs is determined only by the incident power. We measure the power dependence of the TH pulse and determine that the TH-power dependence varies with the focusing diameter. In particular, when the focusing diameter is large, the specific-TH-power dependence appears. The power at which the specific structure begins to appear is determined to be 0.64 GW, which is independent of the focusing diameter and approximately 1/6 of the critical power of the fundamental pulse (3.7 GW). Numerical calculations are performed to verify these experimental results and reproduce the results evidencing that the power at which the specific structure begins to emerge is independent of the focusing diameter. To investigate underlying physics responsible for the specific structure, we calculate the spatial and temporal profiles of the TH pulse and determine that the temporal profile, not the spatial profile, coincides with the power at which its original shape begins to collapse. Furthermore, numerical calculations neglecting the third-order nonlinear effects of fundamental waves, dispersion, and ionization-plasma effects, yield similar results: the time profile of the TH collapses. Numerical calculations also reveal that the cause of the temporal collapse is the interference effects attributed to the third-order nonlinear polarization, which disappears because of the phase mismatch, and higher-order nonlinear terms. Therefore, the experimental results, in which the specific structure begins to emerge for the TH-power dependence, strongly suggest the existence of the  critical power for the TH generation, which is characterized as the time-collapse critical power. Numerical calculations further reveal that the properties of this ``time-collapse critical power” are similar to those of the fundamental-pulse critical power.

The results obtained in this study have broad applicability in the study of nonlinear effects and in unraveling the effects of higher-order nonlinear effects in harmonic generation. In harmonic generation, the lowest-order terms do not disappear, which results in the neglect of the higher-order nonlinear effects. However, the lowest-order nonlinear effects can be canceled by the propagation effects \cite{boyd2020nonlinear}; therefore, the higher-order nonlinear effects can be directly observed. In fact, in harmonic generation in solid, several phenomena have been observed in which higher-order nonlinear effects cannot be ignored \cite{xia2018nonlinear, seres2018non}, and the findings of this study are expected to be useful in such cases.

The critical power has also been theoretically studied because it can be treated analytically \cite{marburger1975self}. Our result extends the concept of the critical power from the fundamental wave to the harmonic generation, opening a new direction for theoretical studies on the critical power for harmonic generation. In particular, what determines this critical power of the TH pulse needs to be explicitly explained in future studies, as well as the explanation of the fundamental pulse described as the balance between the diffraction and the nonlinear Kerr effects. In addition, considering that the nonlinear dynamics of the third harmonic can change at the power smaller than the critical power for the fundamental wave, our results should be useful practically as a benchmark for TH-pulse collapse instead of the common critical power for the fundamental wave.

\begin{acknowledgments}
This work was supported in parts by MEXT Leading Graduates Schools Program,
Advanced Leading Graduate Course for Photon Science (ALPS), MEXT Quantum Leap
Flagship Program (MEXT Q-LEAP) Grant Number JPMXS0118067246, the Center of
Innovation Program funded by the Japan Science and Technology Agency (JST).
\end{acknowledgments}

\appendix
\section{Experimental setup}\label{AppA}
\begin{figure*}
\includegraphics[width=170mm]{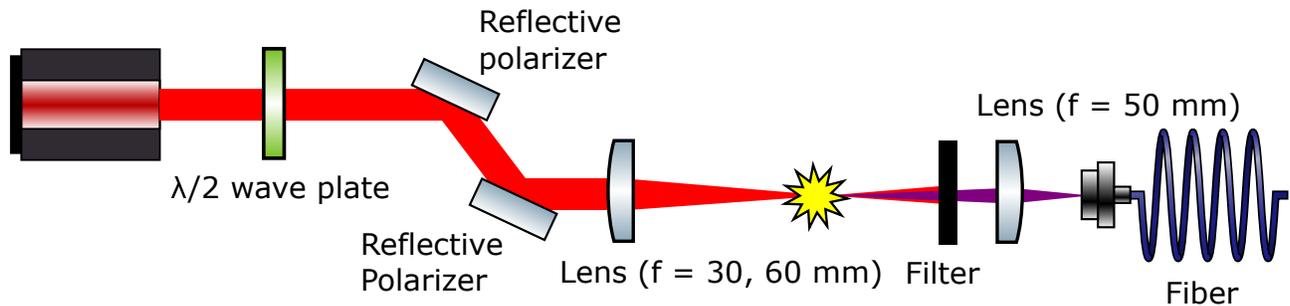}
\caption{\label{sfig1} Experimental setup.}
\end{figure*}

Figure \ref{sfig1} shows a schematic diagram of this experiment. A Ti:sapphire laser source (Astrella, Coherent Inc.) is used as the light source. The laser pulse (center wavelength: 800 nm, pulse width: 45 fs, and repetition rate: 1 kHz) passes through a $\lambda$/2 wave plate and is reflected off a pair of reflective polarizers. Thereafter, the beam is focused into the air by a lens (f = 30, 60 mm). The third-harmonic pulse generated from the air is transmitted through a filter, focused again by a lens (f = 50 mm) onto a fiber, and observed by a spectrometer. The beam profile at the focal point is measured by a CMOS camera (Camera Module V2; Raspberry Pi Foundation) \cite{sakurai2021direct}. 

\section{Numerical simulations}\label{AppB}
Equations (\ref{eq1}, \ref{eq2}, \ref{eq3}, \ref{eq4}, \ref{eq9}) are calculated with the split-step method by Fourier transformation with respect to time and Hankel transformation with respect to space using axial symmetry. The nonlinear terms and electron density are calculated by the fourth-order Runge-Kutta method. As the time coordinate, we use the retarded time $\tau = t - k'_{\omega}z$ \cite{boyd2020nonlinear}. Propagation calculations are performed in the range of $z = -40b$ to $40b$ ($b$: confocal parameter) to accurately include the phase-mismatch effect. The initial condition of the fundamental pulse is expressed as Eq. (\ref{eq6}). The power-dependence graphs in Figures (\ref{fig2}, \ref{sfig4}) show the power of the TH pulse at $z = 40b$, while Figures (\ref{fig3}, \ref{fig4}, \ref{fig5}, \ref{fig6}, \ref{sfig2}, \ref{sfig3}, \ref{sfig5}, and \ref{sfig6}) show the spatial profile of the fundamental and TH pulses at $z = 10b$. The third-order nonlinear susceptibility is half of the values reported in \cite{boyd2020nonlinear} because the nonlinear susceptibility tends to decrease with the pulse widths \cite{thul2020spatially}. Table \ref{stable1} shows the parameters used in the numerical calculations.
\begin{table}
    \begin{center}
        \caption{\label{stable1}Parameters for the numerical simulations.}
        \begin{tabular}{c|c} 
            Parameter & Value \\ \hline
            $\lambda_1$ & 800 (nm) \\
            $n_{\omega}$  & 1.00027505\\
            $n_{3\omega}$ & 1.00029741\\
            $\chi_3$ & 0.85 $\times 10^{-25}$ (m$^2$/V$^2$)\\
            $\Delta k$ & -527 (/m) \\
            $c$ & 299792458 (m/s) \\
            $\hbar$ & 1.05 $\times$ 10$^{-34}$ (Js) \\
            $k'_{\omega}$ & $n_{\omega g}/c, n_{\omega g}$ = 1.00027997 \\
            $k'_{3\omega}$ & $n_{3\omega g}/c, n_{3\omega g}$ = 1.00035508 \\
            $k"_{\omega}$ & 0.21233 $\times$ 10$^{-28}$ (s$^2$/m) \\
            $k"_{3\omega}$ & 1.0079 $\times$ 10$^{-28}$ (s$^2$/m) \\
            $\tau_c$ & 350 (fs) \\
            $N_{mol}$ & 2.67 $\times$ 10$^{19}$ (/cm$^3$) \\
            $N_{O_2}$ & $N_{mol} \times$ 0.21 \\
            $N_{N_2}$ & $N_{mol} \times$ 0.79 \\
            $U_{i O_2}$ & 12 (eV) \\
            $U_{i N_2}$ & 15.6 (eV) \\
            $\sigma_{\omega}$ & 5.5 $\times$ 10$^{-20}$ cm$^2$ \\
            $\sigma_{3\omega}$ & 6.1 $\times$ 10$^{-21}$ cm$^2$ \\
            $\alpha$ & 0.5 (Raman effect) \\
            $\tau$ & 45 (fs)
        \end{tabular}
    \end{center}
\end{table}

\section{Spatial and temporal profiles of the fundamental pulse}\label{AppC}

\begin{figure*}
\includegraphics[width=170mm]{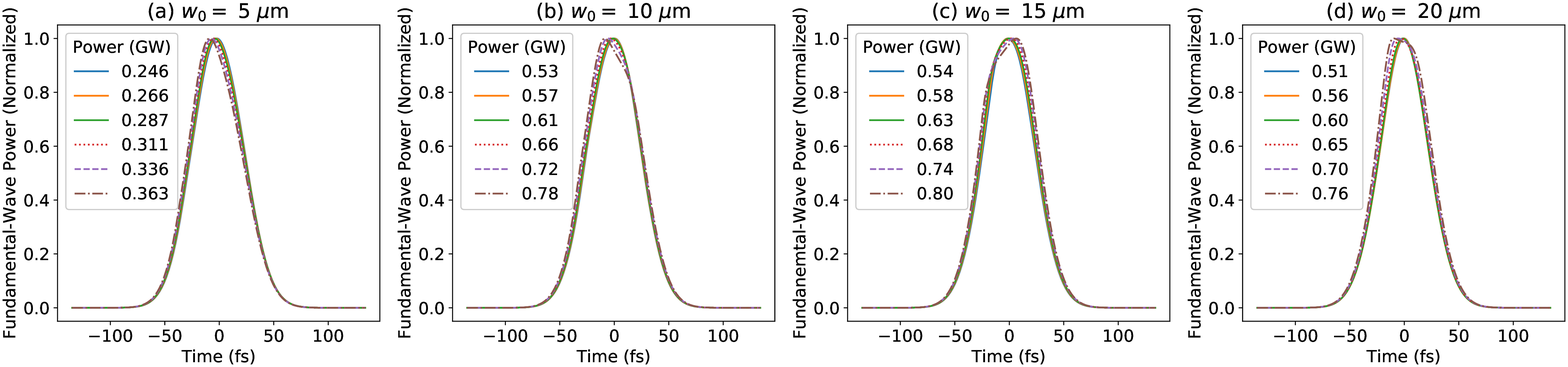}
\caption{\label{sfig2} Calculated fundamental-wave temporal profiles at $r=0$, with the beam waists: (a) $w_0$ = 5 $\mu$m, (b) $w_0$ = 10 $\mu$m, (c) $w_0$ = 15 $\mu$m, and (d) $w_0$ = 20 $\mu$m.}
\end{figure*}

\begin{figure*}
\includegraphics[width=170mm]{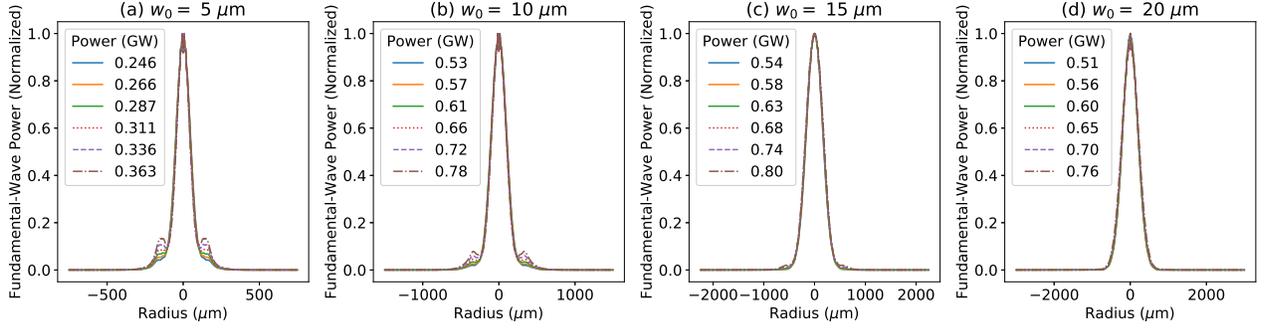}
\caption{\label{sfig3} Calculated fundamental-wave spatial profiles at $t=0$, with the beam waists: (a) $w_0$ = 5 $\mu$m, (b) $w_0$ = 10 $\mu$m, (c) $w_0$ = 15 $\mu$m, and (d) $w_0$ = 20 $\mu$m.}
\end{figure*}

\begin{figure*}
\includegraphics[width=170mm]{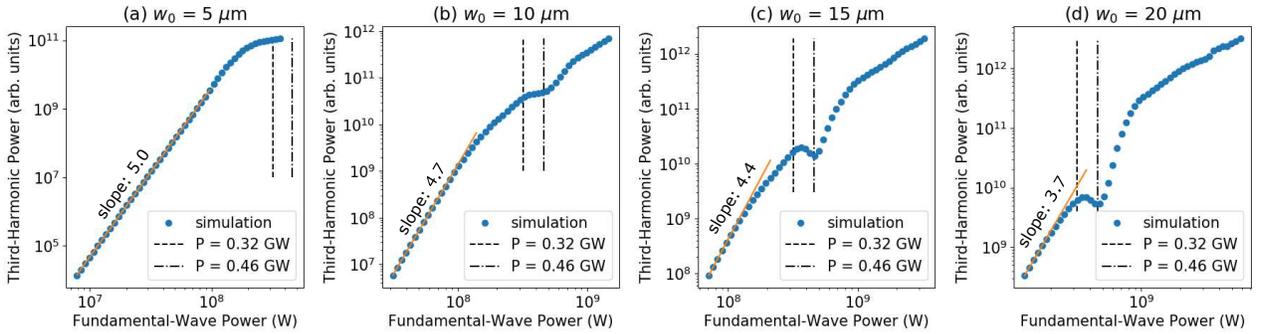}
\caption{\label{sfig4} Calculated TH-power dependences with the beam waists: (a) $w_0$ = 5 $\mu$m, (b) $w_0$ = 10 $\mu$m, (c) $w_0$ = 15 $\mu$m, and (d) $w_0$ = 20 $\mu$m, with the nonlinear susceptibility, $\chi^{(3)}$, being two times larger than that in Figure \ref{fig2}.}
\end{figure*}

\begin{figure}
\includegraphics[width=86mm]{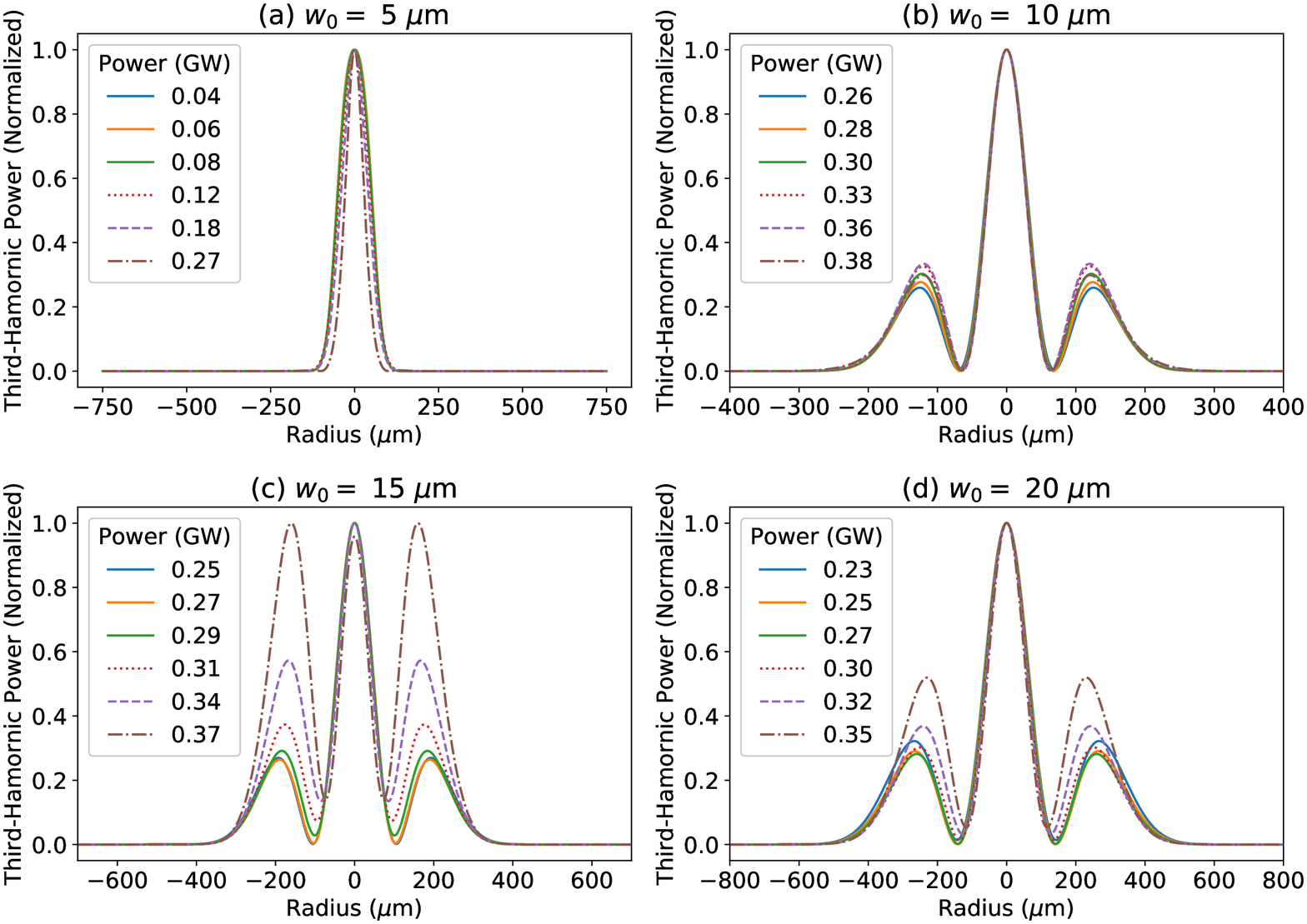}
\caption{\label{sfig5} Calculated TH-spatial profiles at $t=0$, with the nonlinear susceptibility, $\chi^{(3)}$, being two times larger than that in Figure \ref{fig2}. The beam waists are (a) $w_0$ = 5 $\mu$m, (b) $w_0$ = 10 $\mu$m, (c) $w_0$ = 15 $\mu$m, and (d) $w_0$ = 20 $\mu$m.}
\end{figure}

\begin{figure}
\includegraphics[width=86mm]{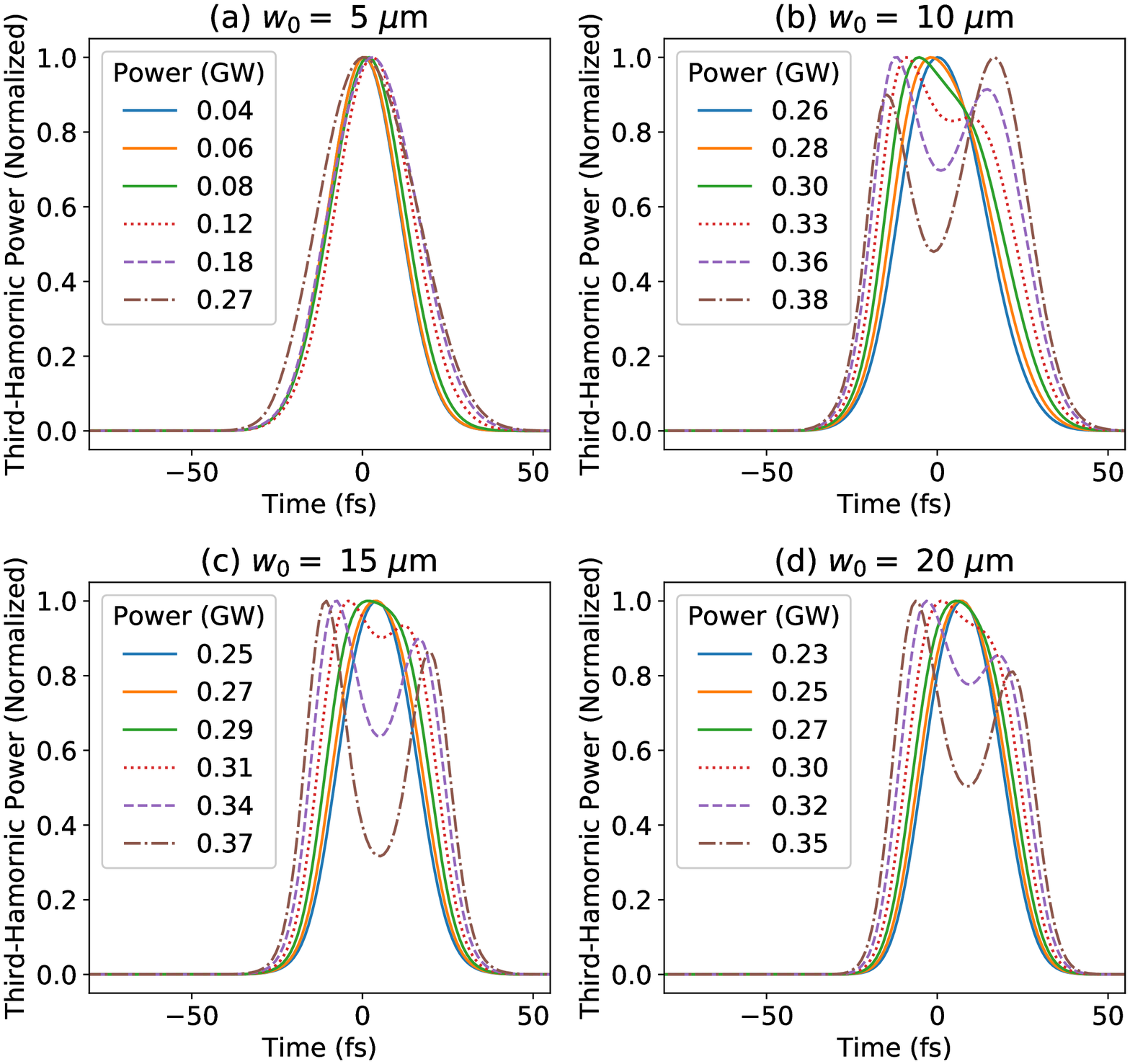}
\caption{\label{sfig6} Calculated TH-temporal profiles at $r=0$, with the nonlinear susceptibility, $\chi^{(3)}$, being two times larger than that in Figure \ref{fig2}. The beam waists are (a) $w_0$ = 5 $\mu$m, (b) $w_0$ = 10 $\mu$m, (c) $w_0$ = 15 $\mu$m, and (d) $w_0$ = 20 $\mu$m.}
\end{figure}

The calculated temporal and spatial profiles of the fundamental pulse are shown in Figures (\ref{sfig2}, \ref{sfig3}), corresponding to those in Figures (\ref{fig2}, \ref{fig3}, and \ref{fig4}). 

Figure \ref{sfig2} shows that the time profile of the fundamental wave does not change from the initial shape independent of the focusing diameter ($w_0$). The same is true for Figure \ref{sfig3}, which shows that the spatial profile of the fundamental wave maintains its original shape, independent of the focusing diameter ($w_0$).


\section{Numerical simulations with the nonlinear susceptibility doubled}\label{AppD}

In the numerical calculations in the main text, the nonlinear susceptibility is $0.85\times 10^{-25}$  m$^2$/V$^2$. Here, the numerical calculations are performed with the doubled value of $1.7\times 10^{-25}$ m$^2$/V$^2$. Figure \ref{sfig4} shows that the critical power of the TH pulse is 0.32 GW, which is half of 0.64 GW in Figure \ref{fig2}. The slope decreases as the focusing diameter increase, which is attributed to the narrowing of the power region to fit the slope. If calculations are conducted using a low incident power, the slope would be 5.0, regardless of the focusing diameter. Figure \ref{sfig5} shows that for (a) $w_0$ = 5 $\mu$m, the spatial profile does not change with the incident power and maintains its original shape. For (b) $w_0$ = 10 $\mu$m, (c) $w_0$ = 15 $\mu$m, and (d) $w_0$ = 20 $\mu$m, the spatial profile deviates from the original shape. These spatial profiles, however, do not change at the power of 0.32 GW. Figure \ref{sfig6} shows the time profiles for the TH pulse. For (a) $w_0$ = 5 $\mu$m, the time profile does not change from the initial shape with the incident power. For (b) $w_0$ = 10 $\mu$m, (c) $w_0$ = 15 $\mu$m, and (d) $w_0$ = 20 $\mu$m, the time profiles begin to collapse when the incident power exceeds 0.3 GW. These calculations are consistent with those in Figures (\ref{fig3}, \ref{fig4}), supporting that the temporal-collapse critical power for the TH pulse is proportional to the inverse of the nonlinear susceptibility, similar to that for the fundamental wave.

\bibliography{apssamp}

\end{document}